\documentclass[%
 reprint,
superscriptaddress,
nofootinbib,
bibnotes,
 amsmath,amssymb,
 aps,
]{revtex4-1}

\usepackage{latexsym,mathrsfs,bbm}
\usepackage{graphicx}
\usepackage{dcolumn}
\usepackage{braket}
\usepackage{gensymb}
\usepackage{bm}
\usepackage{amsmath}
\usepackage{amsthm}
\usepackage{float}
\usepackage{cancel}
\usepackage{mathtools}
\usepackage{color}
\usepackage{csquotes}
\usepackage[utf8]{inputenc}
\usepackage{thmtools}

\declaretheorem[]{theorem}

\usepackage{diagbox}
\usepackage{adjustbox}

\theoremstyle{plain}
\newtheorem{exmp}{Example}
\newtheorem{Cor}{Corollary}

\newcommand{\vecc}{{\rm v}}

\newcommand{\Basis}{\mathcal{B}}  

\newcommand{\JParen}{ {(j)} }

\usepackage{color}
\usepackage[dvipsnames]{xcolor}
\usepackage[normalem]{ulem}

\usepackage{hyperref}

\allowdisplaybreaks

\begin{document}


\title{Conditions tighter than noncommutation needed for nonclassicality}

\author{David R. M. Arvidsson-Shukur}
\affiliation{%
Hitachi Cambridge Laboratory, J. J. Thomson Avenue, CB3 0HE, Cambridge, United Kingdom}
\affiliation{%
 Cavendish Laboratory, Department of Physics, University of Cambridge, Cambridge CB3 0HE, United Kingdom
}%
\affiliation{%
Research Laboratory of Electronics, Massachusetts Institute of Technology, Cambridge, Massachusetts 02139, USA
}%

\author{Jacob Chevalier Drori}
\affiliation{%
 DAMTP, Centre for Mathematical Sciences, University of Cambridge,  Cambridge CB3 0WA, United Kingdom
}%

\author{Nicole Yunger Halpern}
\affiliation{ITAMP, Harvard-Smithsonian Center for Astrophysics, Cambridge, MA 02138, USA}
\affiliation{Department of Physics, Harvard University, Cambridge, MA 02138, USA}
\affiliation{%
Research Laboratory of Electronics, Massachusetts Institute of Technology, Cambridge, Massachusetts 02139, USA
}%
\affiliation{Center for Theoretical Physics, Massachusetts Institute of Technology, Cambridge, Massachusetts 02139, USA}
\affiliation{Joint Center for Quantum Information and Computer Science, NIST and University of Maryland, College Park, MD 20742, USA}
\affiliation{Institute for Physical Science and Technology, University of Maryland, College Park, MD 20742, USA}

\date{\today}

\begin{abstract}
Kirkwood discovered in 1933, and Dirac discovered in 1945, a representation of quantum states that has undergone a renaissance recently. The Kirkwood-Dirac (KD) distribution has been employed to study nonclassicality across quantum physics,  from metrology to chaos to the foundations of quantum theory. The KD distribution is a quasiprobability distribution, a quantum generalization of a probability distribution, which can behave nonclassically by having negative or nonreal elements. Negative KD elements signify quantum information scrambling and potential metrological quantum advantages. Nonreal elements encode measurement disturbance and thermodynamic nonclassicality. KD distributions’ nonclassicality has been believed to follow necessarily from pairwise noncommutation of operators in the distribution's definition. We show that noncommutation does not suffice. We prove sufficient conditions for the KD distribution to be nonclassical (equivalently, necessary conditions for it to be classical). We also quantify the KD nonclassicality achievable under various conditions. This work resolves long-standing questions about nonclassicality and may be used to engineer quantum advantages. \\ MIT-CTP/5278
\end{abstract}

\maketitle


\textit{Introduction.---}Heisenberg's uncertainty principle \cite{Heisenberg27, Kennard27, Landau13} and Bohr's complementarity principle  \cite{Bohr28} power much of the strangeness in quantum mechanics.  The principles codify the  \textit{incompatibility} of simultaneous measurements of certain observables.  Despite incompatibility's essentiality in quantum physics, how  the corresponding nonclassicality is best quantified remains unknown \cite{Designolle19}. Guided by practicality, we use Kirkwood and Dirac’s quasiprobability formalism of quantum mechanics \cite{Kirkwood33, Dirac45}, reviewed below. We prove how operator incompatibility underlies, but does not guarantee, negative and nonreal quasiprobabilities, which signal nonclassical physics under certain circumstances. We then quantify and bound the distribution's nonclassicality.

In classical mechanics, a joint probability-density function $\mathcal{P}(\bm{x}, \bm{p})$ describes a system's  position $\bm{x}$ and momentum $\bm{p}$. In quantum mechanics, observables do not necessarily commute. Representing a state with a joint probability function over observables' eigenvalues is generally impossible \cite{Wigner32, Cohen66, Hudson74, Srinivas75, Hartle04, Allahverdyan15}.

By forfeiting one of Kolmogorov's axioms of joint probability functions \cite{Kolmogorov51}, one can represent quantum mechanics with a probability-like framework. A quantum state can be represented by a quasiprobability function over incompatible observables' eigenvalues. A quasiprobability behaves like a probability but can assume negative and/or nonreal values. Many types of quasiprobability distributions exist. The best-known is the Wigner function, a function of position and momentum \cite{Wigner32, Wootters87, Carmichael13}. The Wigner function (and the related Sudarshan-Glauber P and  Husimi Q representations \cite{Husimi40, Sudarshan63, Glauber63}) are used extensively in quantum optics \cite{Mandel95}, where $\bm{x}$ and $\bm{p}$ are swapped for the electric field's the real and imaginary components. However, in experiments that lack clear analogs of $\bm{x}$ and $\bm{p}$, the Wigner function is less suitable. Furthermore, Wigner-function negativity is neither necessary nor sufficient for nonclassical phenomena: The Einstein-Podolsky-Rosen state \cite{Einstein35} has a positive Wigner function \cite{Revzen05}, and states expressibly classically in the particle-number basis can have negative Wigner representations  \cite{Spekkens08}. 

The Kirkwood-Dirac\footnote{The Kirkwood-Dirac distribution has been called by several names. Its real part is often called  the Terletsky-Margenau-Hill distribution \cite{Terletsky37, Margenau61, Johansen04, Johansen04-2}.} (KD) quasiprobability distribution is a relative of the Wigner function. Kirkwood \cite{Kirkwood33} and Dirac \cite{Dirac45} independently developed the KD distribution to facilitate the application of  probability theory to quantum mechanics. Compared to the Wigner function, the KD distribution possesses  an additional freedom: It can assume nonreal values. Moreover, the KD distribution is straightforwardly defined for discrete systems---even qubits.

The KD distribution has recently illuminated several areas of quantum mechanics. In weak-value amplification \cite{Vaidman88, Duck89, Hosten08, Dixon09}, negative KD quasiprobabilities allow pre- and postselected averages of observables, \textit{weak values}, to lie outside the obervables' eigenspectra, improving signal-to-noise ratios \cite{Steinberg95,  Starling09, Dressel14, Pusey14, Pang14, Pang15, Yunger18, Kunjwal19}. Nonreal KD quasiprobabilities can endow weak values with imaginary components, which encode a measurement's disturbance of a quantum state \cite{Jozsa07, Hofmann11, Dressel12, Monroe20}. Measuring a KD distribution allows for the tomographic  reconstruction of a quantum state \cite{Johansen07, Lundeen11, Lundeen12, Bamber14, Thekkadath16}. In quantum chaos, quantum-information scrambling (the spreading of a local perturbation via many-body entanglement) is quantified with an out-of-time-ordered correlator \cite{Swingle16, Landsman19}. This correlator drops to classically forbidden values when  underlying KD quasiprobabilities assume negative or nonreal values \cite{Yunger17, Yunger18, Yunger18-2, Yunger19, Razieh19}. In quantum metrology, postselection can increase the average amount of information obtained about an unknown parameter per end-of-trial measurement  \cite{ArvShukur17-2, ArvShukur19, ArvShukur19-2, Jenne21}. If the postselection is designed such that a conditional KD distribution contains negative elements, the information-per-final-measurement rate can be nonclassically large.  KD distributions have been used in quantum thermodynamics \cite{Yunger17, Levy19, Lostaglio20}; nonreal KD quasiprobabilities enable
an engine to be unexplainable by any classical (noncontextual) theory  \cite{Lostaglio20}. Finally, the KD distribution has applications to the foundations of quantum mechanics \cite{Griffiths84, Goldstein95, Hartle04, Hofmann11, Hofmann12, Hofmann12-2, Hofmann14, Hofmann15, Hofmann16, Halliwell16, Stacey19}. For example, a KD distribution is related to histories'  weights  in the consistent-histories interpretation of quantum mechanics \cite{Griffiths84, Goldstein95, Hartle04}. Furthermore, nonclassicality of the KD distribution can coincide with the violation of a Leggett-Garg inequality \cite{Hofmann15, Suzuki12}.

Despite  the KD distribution's versatility, many of its properties have not been detailed. A natural first guess is that, if all the operators involved fail to commute with each other pairwise, then the KD distribution contains negative or nonreal quasiprobabilities~\cite{Razieh19}. This, we show below, is a misconception.  Furthermore, little is known about bounds on how much nonclassicality a KD distribution can have.\footnote{
Bounds have been derived on eigenvalues of products of Hermitian operators \cite{Strang62}. Such bounds are of mathematical and fundamental interest. In contrast, we bound KD quasiprobabilities, motivated by the KD distribution's operational significances, as well as by foundational interests.}
An improved understanding of the KD distribution's properties can facilitate the design of diverse experiments that harness the distribution's nonclassicality for quantum advantage.

In this Article, we prove sufficient conditions for the KD distribution to have nonclassically negative and/or nonreal values (Thm. \ref{Th:NClKD}) or, equivalently, necessary conditions for the KD distribution to be classical. We identify cases in which the KD distribution is classical despite pairwise noncommutation between the quantum state and the observables in the distribution's definition. Our results  extend to scenarios where the KD distribution is coarse-grained to account for degeneracies in experiments. Reference \cite{Yunger19} introduced a measure for the KD distribution's nonclassicality. We complement this measure with new ones, suited to more-diverse operational tasks. We also upper-bound these nonclassicality measures 
(Thm. \ref{Th:MaxNCli}). Conditioning the KD distribution, \`{a} la Bayes' theorem, allows KD nonclassicality to exceed the bounds, amplifying  quantum advantages in certain experiments. Finally, we quantify how decoherence reduces KD distributions' nonclassicalities.

\textit{Kirkwood-Dirac distribution.---}We assume that all operators  operate on a Hilbert space with finite dimension $d$. Consider two orthonormal bases, $\{ \ket{a_i} \}$ and $\{ \ket{f_i} \}$. Throughout this article, we regard these bases as eigenbases of observables $\hat{A} = \sum_i a_i \ket{a_i}\bra{a_i}$ and $\hat{F} = \sum_i f_i \ket{f_i}\bra{f_i}$.  In terms of these bases, a state $\hat{\rho}$ can be represented by the KD distribution
\begin{equation}
\label{Eq:KD}
\left\{ q^{\hat{\rho}}_{i,j} \right\} \equiv \left\{  \braket{f_j | a_i} \bra{a_i} \hat{\rho} \ket{ f_j}  \right\} = \left\{ \textrm{Tr}( \hat{\Pi}_{j}^f \hat{\Pi}_{i}^a \hat{\rho}) \right\} ,
\end{equation}
where $ \hat{\Pi}_{i}^a \equiv \ket{a_i}\bra{a_i}$, etc. 
The distribution can be used to calculate expectation values and measurement-outcome probabilities.
$\{ q^{\hat{\rho}}_{i,j} \}$ satisfies some of Kolmogorov's axioms for joint probability distributions \cite{Kolmogorov51}:
\begin{align}
\sum_{i,j} q^{\hat{\rho}}_{i,j}  = 1 , \; \sum_{j} q^{\hat{\rho}}_{i,j}  = p(a_i|\hat{\rho}) , \; \mathrm{and} \;
\sum_{i} q^{\hat{\rho}}_{i,j}  = p(f_j|\hat{\rho}) , \nonumber
\end{align}
where $p(a_i|\hat{\rho})$ and $p(f_j|\hat{\rho})$ denote conditional probabilities.   $q^{\hat{\rho}}_{i,j}$ can be nonclassical by assuming  negative or nonreal values. Nonclassical values are not directly observable but cause effects inferable from sequential measurements \cite{Suzuki16, Yunger18}. If  $\{\ket{a_i}\} = \{\ket{f_j}\}$, the KD distribution reduces to a classical probability distribution: $\{ q^{\hat{\rho}}_{i,j} \} = \{ \braket{f_j | a_i} \bra{a_i} \hat{\rho} \ket{ f_j} \delta_{f_j, a_i} \} = \{ \mathrm{Tr} ( \hat{\Pi}_{i}^a \hat{\rho} ) \delta_{f_j, a_i}  \}$. In classical physics, all observables commute, and every KD distribution equals a probability distribution.


Certain physical processes \cite{Yunger17, Yunger18, Yunger18-2, Yunger19, Razieh19, ArvShukur19-2} motivate the extension of the KD distribution from $2$ to $k$ bases, e.g.,  eigenbases of $k$ observables $\hat{A}^{(1)},\ldots,\hat{A}^{(k)}$. The extended KD distribution is 
\begin{align}
\label{Eq:KDExt}
\left\{ q^{\hat{\rho}}_{i_1,\ldots,i_{k}} \right\} \equiv  \left\{ \textrm{Tr} \left( \hat{\Pi}_{i_{k}}^{a^{(k)}}  \ldots \hat{\Pi}_{{i_1}}^{a^{(1)}} \hat{\rho} \right) \right\} .
\end{align}
A KD distribution's elements serve as the coefficients in an operator expansion of $\hat{\rho}$:
\begin{equation}
\label{Eq:KDDecom}
\hat{\rho} = \sum_{i_1,\ldots,i_k} \frac{\ket{a_{i_1}^{(1)}}\bra{a_{i_{k}}^{(k)}}}{\braket{a_{i_{k}}^{(k)}|a_{i_1}^{(1)}}} q^{\hat{\rho}}_{i_1,\ldots,i_{k}} =\sum_{i, j} \frac{\ket{a_i^{(1)}}\bra{a_j^{(k)}}}{\braket{a_j^{(k)}|a_i^{(1)}}} q^{\hat{\rho}}_{i,j} .
\end{equation}
We define $q^{\hat{\rho}}_{i,j} / \braket{a_j^{(k)}|a_{i}^{(1)}} 
\equiv \bra{a_{i}^{(1)}} \hat{\rho} \ket{ a_j^{(k)}}$ if $\braket{a_j^{(k)}|a_{i}^{(1)}} =0$.

We have shown how to represent a state in terms of eigenbases of Hermitian operators, including measured observables and time-evolution generators. In terms of this representation, physical quantities  can be expressed. Assuming that KD distributions are real and non-negative, one can bound the values attainable  in classical settings. This strategy has been applied  to weak values\footnote{Observables' expectation values equal KD-weighted weak values \cite{Hofmann12-2}.} \cite{Dressel15, Yunger18},  information scrambling \cite{Yunger19, Razieh19}, and the Fisher information \cite{ArvShukur19-2}. Nonclassicality in the KD distribution is a stricter condition than noncommutation, we show, as the former requires the latter but not \textit{vice versa}.


\textit{Requirement for nonclassical quasiprobabilities.---}If any two of $\hat{A}$, $\hat{F}$, and $\hat{\rho}$ commute, they share at least one eigenbasis. When $\hat{A}$ and $\hat{F}$ commute and a shared eigenbasis serves as the $\{ \ket{a_i} \}$ and the $\{ \ket{f_j} \}$ in Eq.~\eqref{Eq:KD}, the KD distribution equals a classical probability distribution. When $\hat{\rho}$ and $\hat{A}$ ($\hat{F}$) commute, it suffices for classicality that a shared eigenbasis serves as $\{ \ket{a_i} \}$  ($\{ \ket{f_j} \}$).  If $[\hat{\rho},\hat{A}] ,[\hat{\rho},\hat{F}] , [\hat{A},\hat{F}] \neq 0$, the KD distribution \textit{may} assume negative or nonreal values \cite{Strang62}. However, noncommutation does not suffice for  KD nonclassicality, as shown in Examples \ref{Ex:NonComPosKDPure} and \ref{Ex:Classical KD which Saturates Bounds}  in App. \ref{App:Ex}. To find a sufficient condition for nonclassicality (equivalently, a necessary condition for classicality), we focus first on (i) pure states $\hat{\rho}$ and (ii) nondegenerate $\hat{A}$ and $\hat{F}$. We then address  degenerate observables and mixed states. 

Let us define four real numbers that reflect incompatibility properties of $\hat{\rho}$,  $\hat{A}$, and $\hat{F}$. In the pure case, $\hat{\rho}  = \ket{\Psi}\bra{\Psi}$. Let $\mathcal{V}_A \equiv \{ \ket{a_i} \}$ and $\mathcal{V}_F \equiv \{ \ket{f_j} \}$ denote the eigenbases of the nondegenerate  $\hat{A}$ and $\hat{F}$, respectively. These eigenbases are unique up to phases. Define  as $N_A$ ($N_F$)  the number of $\mathcal{V}_A$ ($\mathcal{V}_F$) vectors whose overlaps with $\ket{\Psi}$ are nonzero:
\begin{align}
\label{Eq:Na}
N_A & \equiv || \{ \ket{a_i} \in \mathcal{V}_A : \braket{a_i|\Psi} \neq 0 \} || , \; \mathrm{and} \\
\label{Eq:Nf}
N_F & \equiv || \{ \ket{f_j} \in \mathcal{V}_F : \braket{f_j|\Psi} \neq 0 \} || .
\end{align}
$|| \cdot ||$ denotes a set's cardinality. We denote by $n_\parallel$ ($\bar{n}_\parallel$) 
the number of $\ket{a_i}$ that are
(i) parallel to vectors $\ket{f_j}$ and
(ii) nonorthogonal (orthogonal) to $\ket{\Psi}$. 

\begin{theorem}[Sufficient conditions for Kirkwood-Dirac nonclassicality]
\label{Th:NClKD}
Suppose that $\hat{\rho}$ is pure and that $\hat{A}$ and $\hat{F}$ are nondegenerate. If $2 N_A + 2 N_F > 3d+n_{\parallel}-3\bar{n}_{\parallel}$,  then the Kirkwood-Dirac distribution contains negative or nonreal values.
\end{theorem}

We prove the theorem by the contrapositive: 
Assuming a classical KD distribution, we deduce constraints on the unitary matrix with entries $\braket{a_i | f_j}$. These constraints imply a condition on $N_A$, $N_F$, $d$, $n_{\parallel}$, and $\bar{n}_{\parallel}$
that is necessary for classicality of the KD distribution. 
A violation of this condition suffices for KD nonclassicality. The full proof appears in App. \ref{App:Theo1Proof}

Theorem \ref{Th:NClKD} implies a simple condition sufficient for KD nonclassicality:

\begin{Cor}
\label{Cor:KD0}
	If the KD distribution lacks zero-valued quasiprobabilities, $ \{ q^{\hat{\rho}}_{i,j} \}$ is nonclassical.
\end{Cor}

\noindent \textit{Proof:} If all $q^{\hat{\rho}}_{i,j} \neq 0$, then $\ket{a_i} \not\parallel  \ket{f_j}$,\footnote{ 
Since $\mathcal{V}_A$ and $\mathcal{V}_F$ are orthonormal sets, if some  $\ket{a_i} \parallel  \ket{f_j}$, then some other  $\ket{a_{i^{\prime}}} \perp \ket{f_j}$. By Eq. \eqref{Eq:KD}, $q^{\hat{\rho}}_{i^{\prime},j} = 0$. } 
and $\braket{a_i | \Psi}, \braket{f_j | \Psi} \neq 0$,  for all $i, \, j$.
So $n_\parallel=\bar{n}_\parallel=0$, and $N_A=N_F=d$, satisfying the nonclassicality condition of Thm.~\ref{Th:NClKD}. $\square$

Three more extensions of Thm.~\ref{Th:NClKD} merit mention. First, if $\hat{A}$ and $\hat{F}$ are degenerate,
one can construct KD distributions by coarse-graining over the degeneracies.
These coarse-grained distributions can signal nonclassical physics in quantum chaos \cite{Yunger18, Yunger18-2, Yunger19, Razieh19} and metrology \cite{ArvShukur19-2}.
In App.~\ref{App:CoarseKDExt}, we prove sufficient conditions for these distributions to be nonclassical.

Second, every KD distribution $ \{ q^{\hat{\rho}}_{i_1,i_k} \} $ follows from  marginalizing an extended distribution $ \{ q^{\hat{\rho}}_{i_1,\ldots,i_k} \} $ 
[Eq.~\eqref{Eq:KDExt}]
over the indices $i_2,\ldots,i_{k-1}$ \cite{Yunger17, Yunger18, Yunger18-2, Yunger19, Razieh19, ArvShukur19-2}. If any marginalized $\{q^{\hat{\rho}}_{i_{\alpha}, i_{\beta}} \}$ satisfies the nonclassicality condition in Thm. \ref{Th:NClKD}, every fine-graining $\{ q^{\hat{\rho}}_{i_1,\ldots,i_k} \} $ is nonclassical.


Third, we prove further properties of the real and imaginary components of $q^{\hat{\rho}}_{i,j}$ in App. \ref{App:ReImKD}. These properties can be used, e.g., to tailor states $\hat{\rho}$ to achieve nonclassical results 
in experiments that involve observables $\hat{A}$ and $\hat{F}$. A similar strategy is being applied in a photonic experiment to observe how KD negativity benefits parameter estimation \cite{Lupu20}.


\textit{Nonclassicality measures.---}How much nonclassicality can a KD distribution have? 
We review an existing nonclassicality measure, define measures suited to 
more operational tasks, and upper-bound the measures.

Every KD distribution's elements sum to unity. Negative and nonreal entries are nonclassical.  Gonz\'{a}lez  Alonso \textit{et al.} thus quantified \cite{Yunger19}  
KD distributions'  nonclassicality, in the context of scrambling, with
\begin{equation}
\label{Eq:AggNonClas}
\mathcal{N} \left( \{ q^{\hat{\rho}}_{i_1, \ldots, i_k} \} \right) \equiv - 1 +  \sum_{i_1, \ldots, i_k} \big| q^{\hat{\rho}}_{i_1, \ldots, i_k} \big| .
\end{equation} 
$\mathcal{N} ( \{ q^{\hat{\rho}}_{i_1, \ldots, i_k} \} ) = 0$ when $ \{ q^{\hat{\rho}}_{i_1, \ldots, i_k} \}$ is real and non-negative.
We upper-bound the measure generally in terms of the Hilbert-space dimensionality, $d$. 

\begin{theorem}[Maximum Kirkwood-Dirac nonclassicality]
\label{Th:MaxNCli}
The maximum nonclassicality $\mathcal{N} ( \{ q^{\hat{\rho}}_{i_1, \ldots, i_k} \})$ 
of any Kirkwood-Dirac distribution 
$ \{ q^{\hat{\rho}}_{i_1, \ldots, i_k} \}$ is
\begin{equation}
\label{Eq:AggNonClasMax}
\max_{\hat{\rho}, \hat{A}^{(1)},\ldots,\hat{A}^{(k)}} \Big\{ \mathcal{N} \left( \{ q^{ \hat{\rho} }_{i_1, \ldots, i_k} \} \right) \Big\}= d^{(k-1)/2}-1 .
\end{equation}
The maximum is achieved if and only if two conditions are met simultaneously:
(i) The operators $\hat{A}^{(i)}$ and $\hat{A}^{(i+1)}$  have mutually unbiased eigenbases\footnote{
Bases $\mathcal{A} \equiv \{ \ket{\alpha_j} \}$ and $\mathcal{B} \equiv\{ \ket{\beta_k} \}$ are 
	\emph{mutually unbiased}
	if preparing any $\mathcal{A}$ element and measuring $\mathcal{B}$ yields a totally unpredictable outcome: $|\braket{\alpha_j | \beta_k}| = 1/\sqrt{d}$ for all $j,k$.} 
(MUBs) for each $i = 1,\ldots,k-1$.
(ii) $\hat{\rho} = \ket{\Psi} \bra{\Psi}$, where $\ket{\Psi}$ has equal overlaps with all the eigenvectors of $\hat{A}^{(1)}$ and $\hat{A}^{(k)}$. 
\end{theorem}
The proof of Thm. \ref{Th:MaxNCli} appears in App.  \ref{App:Theo2Proof}.

At least one triplet of MUBs exists for every $d \geq 2$  \cite{Durt10}. 
We can therefore construct a  
$ \left\{ q^{\ket{\Psi}\bra{\Psi}}_{i_1, \ldots, i_k} \right\}$ 
that maximizes $\mathcal{N}$:
Let $\ket{\Psi}$ be an element of the triplet's first MUB.
Let $\ket{a^{(k)}_{i_k}}$ be 
the $i_k^{\mathrm{th}}$ element of 
the second (third) MUB 
if $k$ is even (odd).

The measure \eqref{Eq:AggNonClas} is useful in the context of 
chaos, where negative and nonreal KD quasiprobabilities signal scrambling \cite{Razieh19}. But negative and nonreal values do not always enjoy equal footing: Only negative KD quasiprobabilities enable a metrologist to garner a nonclassically high Fisher information \cite{ArvShukur19-2}. In contrast, nonreal KD quasiprobabilities lie behind weak values' imaginary components, which encode measurement disturbance \cite{Steinberg95, Dressel12}. We therefore quantify the aggregated negativity and nonreality, respectively:
\begin{align}
\label{Eq:AggNonClasRe}
\mathcal{N}^{\Re^{-}} \left( \{ q^{\hat{\rho}}_{i_1, \ldots, i_k} \} \right) & :=   -1 +  \sum_{i_1, \ldots, i_k}     \big|  \Re(q^{\hat{\rho}}_{i_1, \ldots, i_k})  \big|    , \; \mathrm{and}\\
\label{Eq:AggNonClasIm}
\mathcal{N}^{\Im} \left( \{ q^{\hat{\rho}}_{i_1, \ldots, i_k} \} \right) & :=   \sum_{i_1, \ldots, i_k}  \big|  \Im(q^{\hat{\rho}}_{i_1, \ldots, i_k})  \big|   .
\end{align}

$\mathcal{N}^{\Re^{-}} \leq \mathcal{N}$ by definition, and $0 \leq \mathcal{N}^{\Im} < \mathcal{N} +1$.
If all the nonclassical $q^{\hat{\rho}}_{i_1, \ldots, i_k}$ are real negative numbers,
$\mathcal{N}^{\Re^{-}} = \mathcal{N}$. Given the importance of $\mathcal{N}^{\Re^{-}}$ to quantum metrology and weak-value amplification, a crucial question is: When can $\mathcal{N}^{\Re^{-}} = \max\{ \mathcal{N} \}$? A complete answer requires further advances in the field of MUBs. 
Nevertheless, for every $d$ in which a triplet of real MUBs exists,\footnote{For our purposes, a real MUB is an MUB whose vectors can be expressed, relative to a fixed basis, as columns of real numbers. Appendix \ref{App:RealMUBs} reconciles this definition with the conventional definition.} $ \max\{ \mathcal{N}^{\Re^{-}} \} = \max\{ \mathcal{N} \}$. 
The number of real MUBs in a space of a general dimensionality $d$ is unknown. 
The smallest space with a triplet of real MUBs has $d=4$ \cite{Boykin05}.
We construct an example in which $d=4$ and
$\max\{ \mathcal{N}^{\Re^{-}} \} = \max\{ \mathcal{N} \}$
in Ex. \ref{Ex:Saturated Nonclassicality} of App. \ref{App:Ex}.  
In $d=2$, the Pauli bases form a triplet of MUBs. When $k=2$ and the Pauli bases are used to maximize $\mathcal{N}$, all nonclassicality manifests as nonreal quasiprobabilities without negative real  components (App. \ref{App:Ex}, Ex. \ref{Ex:PauliKD}).

\textit{Amplifying nonclassicality via postselection.---}As aforementioned,  negative KD quasiprobabilities underlie quantum advantages in weak-value amplification and postselected quantum metrology. The reason is, the protocols involve postselection. Classical postselection, or conditioning, obeys Bayes' theorem, $p(a|b) = p(b|a) p(a) / p(b)$. The KD distribution satisfies an analog of Bayes' theorem \cite{Johansen04, Johansen07,  Yunger18}: Suppose that a state represented by $ \{ q^{\hat{\rho}}_{i_1, \ldots, i_k} \}$ undergoes a measurement $\{\hat{F}_k, \hat{1} - \hat{F}_k \}$, where $\hat{F}_k \equiv \sum_{i_k \, : \, \ket{f_{i_k}} \in \mathcal{F}_k}
\ket{f_{i_k}} \bra{f_{i_k}}$ for some set $\mathcal{F}_k$. Conditioned on the outcome's corresponding to $\hat{F}_k$,  the  KD quasiprobabilities are  
\begin{align}
\label{Eq:CondKD}
&   \frac{\sum_{i_k \, : \, \ket{f_{i_k}} \in \mathcal{F}_k }q^{\hat{\rho}}_{i_1, \ldots, i_k} }{p( F_k | \hat{\rho})}   , \;\;\; \textrm{where} \\
&p( F_k  | \hat{\rho})  \equiv \sum_{
\substack{i_1, \ldots, i_{k-1}, \\ i_k \; : \, \ket{f_{i_k}} \in \mathcal{F}_k } }q^{\hat{\rho}}_{i_1, \ldots, i_k} 
= \mathrm{Tr}( \hat{F}_k \hat{\rho}).
\end{align}
The form of $q^{\hat{\rho}}_{i_1, \ldots, i_k}$ [Eq. \eqref{Eq:KD}] implies that, for every unconditioned KD distribution, $0\leq |q^{\hat{\rho}}_{i_1, \ldots, i_k}| \leq 1$. If $ \{ q^{\hat{\rho}}_{i_1, \ldots, i_k} \}$ lacks nonclassical values, also the conditional KD quasiprobabilities \eqref{Eq:CondKD} lie between $0$ and $1$. However, if $q^{\hat{\rho}}_{i_1, \ldots, i_k}$ contains negative values, the numerator in Eq. \eqref{Eq:CondKD} can have a greater magnitude than the denominator. The conditional quasiprobability can be made arbitrarily large \cite{ArvShukur19-2}. So can, consequently, the corresponding $\mathcal{N}$, $\mathcal{N}^{\Re^{-}} $, and $\mathcal{N}^{\Im}$. This KD nonclassicality can lead to metrological capabilities infinitely greater than those achievable classically [sometimes at a cost of low postselection probabilities $p( F_k  | \hat{\rho})$]~\cite{Vaidman88, Duck89, ArvShukur19-2}.

\textit{Mixed states.---}We have focused on  pure-state KD distributions, but every experiment involves  decoherence. How does decoherence affect KD nonclassicality?  Let $\hat{\rho} = \sum_n p_n  \hat{\rho}_n$, where $\hat{\rho}_n \equiv \ket{\Psi_n} \bra{\Psi_n}$ and $p_n$ denotes a probability. $\hat{\rho}$ can be represented by the KD distribution$ \{ q^{\hat{\rho}}_{i,j} \}=\{ \sum_n p_n q^{\hat{\rho}_n}_{i,j} \}$. By convexity, the nonclassical $q^{\hat{\rho}}_{i,j}$
have magnitudes no greater than the magnitudes of the nonclassical components of the most nonclassical $ \{ q^{\hat{\rho}_n}_{i,j} \}$:
Mixing dilutes the nonclassicality.  For example, the KD distributions for the pure states  $\hat{\rho}_+ = \ket{+}\bra{+}$ and $\hat{\rho}_- =\ket{-}\bra{-}$   with respect to the  bases $\{  \ket{a} \} = \{  \ket{0}, \ket{1} \}  $ and $\{  \ket{f} \} = \{  \cos{(\pi/3)}\ket{0} + \sin{(\pi/3)} \ket{1},  -\sin{(\pi/3)}\ket{0} + \cos{(\pi/3)} \ket{1} \}$ are  nonclassical. But the  distribution for $\hat{\rho} = \frac{2}{3}\hat{\rho}_++\frac{1}{3}\hat{\rho}_-$ is  classical.\footnote{
$\ket{+}$ ($\ket{-}$) and $\ket{0}$ ($\ket{1}$) denote the $+1$ ($-1$) eigenvectors of the Pauli-$x$ and Pauli-$z$ operators, respectively.} Decoherence obscures the incompatible eigenbases' nonclassicality.

In another example, consider depolarizing a pure state $ \hat{\rho}_0 $: $
\hat{\rho}^{\prime} \equiv p \hat{\rho}_0  + (1-p) \hat{1} / d $.
The KD distribution of $\hat{\rho}^{\prime}$ has elements
\begin{align}
\label{Eq:Depol}
   q^{\hat{\rho}^{\prime}}_{i,j}   = p  \, q^{\hat{\rho}_0}_{i,j} + \frac{1-p}{d}  \, |   \braket{f_j | a_i} |^2 . 
\end{align}
If $p$ is small enough (e.g., if $p=0$), the depolarizing channel eliminates the KD distribution's negative components. By the triangle inequality, $\mathcal{N} ( \{ q^{\hat{\rho}^{\prime}}_{i,j} \} ) \leq  p \mathcal{N}( \{ q^{\hat{\rho}_0}_{i,j} \} ) $, and $\mathcal{N}^{\Re^{-}} ( \{  q^{\hat{\rho}^{\prime}}_{i,j} \} ) \leq  p \mathcal{N}^{\Re^{-}}( \{ q^{\hat{\rho}_0}_{i,j} \} ) $. Each imaginary component is reduced by a factor of $p$: $ \mathcal{N}^{\Im}( \{ q^{\hat{\rho}^{\prime}}_{i,j} \} ) = p \mathcal{N}^{\Im}( \{ q^{\hat{\rho}_0}_{i,j} \} ) $.  $ \mathcal{N}^{\Im}( \{ q^{\hat{\rho}^{\prime}}_{i,j} \} ) $ can resist decoherence more than $\mathcal{N}^{\Re^{-}} ( \{ q^{\hat{\rho}^{\prime}}_{i,j} \} )$: Only when the state decoheres fully ($p=0$) do all the imaginary components disappear. The negative components disappear 
when the decoherence surpasses a finite threshold.

\textit{Discussion.---}Benefits of using the KD distribution include the ability to prove classical bounds on physical quantities by assuming real, non-negative distributions. The key to applying the KD distribution fruitfully  is to construct the distribution operationally. The bases and their ordering should reflect properties of the experiment (e.g.,  \cite{Yunger18, ArvShukur19-2, Razieh19, Lostaglio20, Lupu20}). Similarly, experimental context dictates when extending the KD distribution facilitates analyses~\cite{Yunger17, Yunger18, Yunger18-2, Yunger19, ArvShukur19-2, Razieh19}.

Our work provides a methodology for calculating whether an input state and subsequent operations may generate nonclassical physics in a range of experiments. Furthermore, our work provides a mathematical toolkit for constructing quantum-enhanced experiments. We have shown that noncommutation does not suffice for achieving nonclassical KD distributions and associated quantum advantages. Instead, KD negativity and nonreality emerge as sharper nonclassicality criteria than noncommutation for diverse tasks.

\medskip

\textit{Acknowledgements.---}The authors would like to thank Crispin Barnes, Stephan de Bi\`{e}vre, Nicolas Delfosse, Giacomo De Palma, and Justin Dressel  for useful discussions. D.R.M.A.-S. was supported by the EPSRC, Lars Hierta's Memorial Foundation, and Girton College. N.Y.H. was supported by an NSF grant for the Institute for Theoretical Atomic, Molecular, and Optical Physics at Harvard University and the Smithsonian Astrophysical Observatory and by the MIT CTP administratively.


 \appendix
\onecolumngrid

   \section*{Supplementary Material} 
   
   \section{Example KD distributions} \label{App:Ex} 
 
    \begin{exmp}[Classical KD distribution for pairwise-noncommuting $\hat{A}$, $\hat{F}$, and pure $\hat{\rho}$] \label{Ex:NonComPosKDPure} 
Consider a two-qubit system.  As before, $\ket{+}$ ($\ket{-}$) and $\ket{0}$ ($\ket{1}$) are the $+1$ ($-1$) eigenvectors of the Pauli-$x$ and Pauli-$z$ operators, respectively. We choose  $\hat{A}$ and $\hat{F}$ such that  $\{ \ket{a_i} \} = \{ \ket{0}\ket{0}, \ket{0}\ket{1}, \ket{1}\ket{0}, \ket{1}\ket{1} \}$ and $\{ \ket{f_j} \} = \{ \ket{0}\ket{+}, \ket{0}\ket{-}, \ket{1}\ket{0}, \ket{1}\ket{1} \}$. For example, if each observable has the eigenvalues $-2$, $-1$, $1$, and $2$,
\begin{equation}
 \hat{A} \to
\begin{pmatrix}
   -2 & 0 & 0 & 0 \\
    0 & -1 & 0 & 0 \\
     0 & 0 & 1 & 0\\
    0 & 0 & 0 & 2
 \end{pmatrix} , \; \mathrm{and} \; 
 \hat{F} \to
 \begin{pmatrix}
   -\frac{3}{2} & -\frac{1}{2} & 0 & 0 \\[2pt]
    -\frac{1}{2} & -\frac{3}{2} & 0 & 0 \\[2pt]
     0 & 0 & 1 & 0\\[2pt]
    0 & 0 & 0 & 2
 \end{pmatrix} .
\end{equation}
We set $\hat{\rho} = \ket{\Psi} \bra{\Psi}$, where $\ket{\Psi} = \ket{1}\ket{+}$:
\begin{equation}
 \hat{\rho} \to
\begin{pmatrix}
   0 & 0 & 0 & 0 \\[2pt]
    0 & 0 & 0 & 0 \\[2pt]
     0 & 0 & \frac{1}{2} & \frac{1}{2}\\[2pt]
    0 & 0 & \frac{1}{2} & \frac{1}{2}
 \end{pmatrix} .
\end{equation}
$\hat{A}$, $\hat{F}$ and $\hat{\rho}$ fail to commute pairwise: $[\hat{A}, \hat{F}], \, [\hat{\rho}, \hat{A}], \, [\hat{\rho}, \hat{F}]  \neq 0 $. However, the KD distribution (Table \ref{Tab:NonComPosKDPure}) is real and non-negative.

\begin{table}[h]
\caption{\label{Tab:NonComPosKDPure}%
The KD distribution of Ex. \ref{Ex:NonComPosKDPure}.
}
\begin{center}
\begin{tabular}{c|cccc}
\textrm{\backslashbox{$\ket{f_j}$}{$\ket{a_i}$} } &
\textrm{$\ket{0}\ket{0}$} &
\textrm{$\ket{0}\ket{1}$}&
\textrm{$\ket{1}\ket{0}$}&
\textrm{$\ket{1}\ket{1}$}\\
\hline \hline 
$\ket{0}\ket{+}$ & $0$ & $0$ & $0$ & $0$ \\
$\ket{0}\ket{-}$ & $0$ & $0$ & $0$ & $0$
 \\
$\ket{1}\ket{0}$ & $0$ & $0$ & $\frac{1}{2}$& $0$
 \\
$\ket{1}\ket{1}$ & $0$ & $0$ & $0$ & $\frac{1}{2}$
\end{tabular}
\end{center}
\end{table}

\noindent Since this KD distribution is classical, Thm. \ref{Th:NClKD} implies that $2N_A+2N_F \leq 3d+n_\parallel-3\bar{n}_\parallel$. Indeed, $N_A=N_F=2$, $d=4$, $n_\parallel =2$, and $ \bar{n}_\parallel=0$; so the inequality reads $8\leq 14$.
\\

\end{exmp}

\begin{exmp}[Classical KD distribution that saturates Ineq.  \eqref{Eq:ReducedBound}] \label{Ex:Classical KD which Saturates Bounds} 

Consider a $4$-dimensional Hilbert space with an orthonormal basis $\{\ket{0},\ket{1},\ket{2},\ket{3}\}$. Suppose that $\hat{A}$ and $\hat{F}$ have eigenbases $\{ \ket{a_i} \} = \{\ket{0},\ket{1},\ket{2},\ket{3}\}$ and $\{ \ket{f_j} \} = \{ \frac{\ket{0}+\ket{1}}{\sqrt{2}},\frac{\ket{0}-\ket{1}}{\sqrt{2}},\frac{\ket{2}+\ket{3}}{\sqrt{2}},\frac{\ket{2}-\ket{3}}{\sqrt{2}} \}$. Let $\hat{\rho} = \ket{\Psi}\bra{\Psi}$, where $\ket{\Psi}=\frac{\ket{0}+\ket{1}+\ket{2}+\ket{3}}{2}$. The KD distribution, presented in Table \ref{Tab:Classical KD which Saturates Bounds}, is real and non-negative.

\begin{table}[h]
	\caption{\label{Tab:Classical KD which Saturates Bounds}%
		The KD distribution of Ex. \ref{Ex:Classical KD which Saturates Bounds}.
	}
	\begin{center}
		\begin{tabular}{c|cccc}
			\textrm{\backslashbox{$\ket{f_j}$}{$\ket{a_i}$} } &
			\textrm{$\ket{0}$} &
			\textrm{$\ket{1}$}&
			\textrm{$\ket{2}$}&
			\textrm{$\ket{3}$}\\
			\hline \hline 
			$\frac{\ket{0}+\ket{1}}{\sqrt{2}}$ & $\frac{1}{4}$ & $\frac{1}{4}$  & $0$ & $0$ \\
			$\frac{\ket{0}-\ket{1}}{\sqrt{2}}$ & $0$ & $0$ & $0$ & $0$
			\\
			$\frac{\ket{2}+\ket{3}}{\sqrt{2}}$ & $0$ & $0$ & $\frac{1}{4}$  & $\frac{1}{4}$ 
			\\
			$\frac{\ket{2}-\ket{3}}{\sqrt{2}}$ & $0$ & $0$ & $0$ & $0$
		\end{tabular}
	\end{center}
\end{table}

\noindent In this example, $N_A=4$, $N_F=2$, $d=4$, and $n_\parallel=\bar{n}_\parallel=0$.
Hence, $2N_A+2N_F=12=3d+n_\parallel-3\bar{n}_\parallel$:
The classical inequality $2N_A+2N_F\leq 3d+n_\parallel-3\bar{n}_\parallel$  obtained from  Thm. \ref{Th:NClKD} is saturated.

\end{exmp}

\begin{exmp}[Real nonclassical KD distribution that achieves the maximum in Thm. \ref{Th:MaxNCli}] \label{Ex:Saturated Nonclassicality}

Suppose that $\hat{A}$ and $\hat{F}$ act on a two-qubit Hilbert space and have eigenbases $\{ \ket{a_i} \} = \{ \ket{0}\ket{0}, \ket{0}\ket{1}, \ket{1}\ket{0}, \ket{1}\ket{1} \}$ and $\{ \ket{f_j} \} = \{ \ket{+}\ket{+}, \ket{-}\ket{+}, \ket{+}\ket{-}, \ket{-}\ket{-} \}$. $\{ \ket{a_i} \}$ and $\{ \ket{f_j} \}$ form a pair of MUBs. Let $\hat{\rho} = \ket{\Psi}\bra{\Psi}$, where $\ket{\Psi}=(\ket{0}\ket{0}+\ket{0}\ket{1}+\ket{1}\ket{0}-\ket{1}\ket{1})/2$. The overlaps $|\braket{ \Psi | a_i }| = |\braket{ \Psi | f_j }| = |\braket{ a_i | f_j }| = \frac{1}{2}$ for all $i,j$. The resulting KD distribution is given in Table \ref{Tab:Saturated Nonclassicality}.

\begin{table}[h]
	\caption{\label{Tab:Saturated Nonclassicality}
		The KD distribution of Ex. \ref{Ex:Saturated Nonclassicality}.
	}
	\begin{center}
		\begin{tabular}{c|cccc}
			\textrm{\backslashbox{$\ket{f_j}$}{$\ket{a_i}$} } &
			\textrm{$\ket{0}\ket{0}$} &
			\textrm{$\ket{0}\ket{1}$}&
			\textrm{$\ket{1}\ket{0}$}&
			\textrm{$\ket{1}\ket{1}$}\\
			\hline \hline 
			$\ket{+}\ket{+}$ & $\frac{1}{8}$ & $\frac{1}{8}$ & $\frac{1}{8}$ & $-\frac{1}{8}$ \\
			$\ket{-}\ket{+}$ & $\frac{1}{8}$ & $\frac{1}{8}$ & $-\frac{1}{8}$ & $\frac{1}{8}$
			\\
			$\ket{+}\ket{-}$ & $\frac{1}{8}$ & $-\frac{1}{8}$ & $\frac{1}{8}$ & $\frac{1}{8}$
			\\
			$\ket{-}\ket{-}$ & $-\frac{1}{8}$ & $\frac{1}{8}$ & $\frac{1}{8}$ & $\frac{1}{8}$
		\end{tabular}
	\end{center}
\end{table}
This KD distribution is nonclassical. Furthermore, $\mathcal{N} \left( \left\{q^{\ket{\Psi}\bra{\Psi}}_{i,j}  \right\} \right) =1$ saturates the inequality $\mathcal{N} \left( \left\{ q^{\ket{\Psi}\bra{\Psi}}_{i_1, \ldots, i_k} \right\} \right) \leq d^{(k-1)/2}-1$  in Thm. \ref{Th:MaxNCli}, for  $k=2$ and $d=4$. As the KD distribution is real, it  saturates also $\mathcal{N}^{\Re^{-}} \leq \mathcal{N}$.

\end{exmp}

\begin{exmp}[Nonclassical KD distribution for Pauli operators] \label{Ex:PauliKD}

Let $\hat{A} = \hat{\sigma}_z$, $\hat{F} = \hat{\sigma}_x$, and $\hat{\rho} = \ket{\Psi} \bra{\Psi}$, where $(  \ket{0} + i \ket{1}) /\sqrt{2}$ (the $+1$ eigenstate of $\hat{\sigma}_y$).  The resulting KD distribution is given in Table \ref{Tab:PauliKD}.

\begin{table}[h]
	\caption{\label{Tab:PauliKD}
		The KD distribution of Ex. \ref{Ex:PauliKD}.
	}
	\begin{center}
		\begin{tabular}{c|cccc}
			\textrm{\backslashbox{$\ket{f_j}$}{$\ket{a_i}$} } &
			\textrm{$\ket{0}$} &
			\textrm{$\ket{1}$}\\
			\hline \hline 
			$\ket{+}$ & $(1-i)/4$ & $(1+i)/4$  
			\\
			$\ket{-}$ & $(1+i)/4$ & $(1-i)/4$
		\end{tabular}
	\end{center}
\end{table}

This KD distribution is nonclassical. Furthermore,  $\mathcal{N} \left( \left\{q^{\ket{\Psi}\bra{\Psi}}_{i,j}  \right\} \right)=\sqrt{2}-1$ saturates the inequality $\mathcal{N} \left( \left\{q^{\ket{\Psi}\bra{\Psi}}_{i,j}  \right\} \right) \leq d^{(k-1)/2}-1$  in Thm. \ref{Th:MaxNCli}, for $k=2$ and $d=2$. The KD distribution is non-negative, so  $\mathcal{N}^{\Re^{-}}\left( \left\{q^{\ket{\Psi}\bra{\Psi}}_{i,j}  \right\} \right) = 0$. All the nonclassicality lies in the imaginary components of $\left\{q^{\ket{\Psi}\bra{\Psi}}_{i,j}  \right\}$:  $\mathcal{N}^{\Im} \left( \left\{q^{\ket{\Psi}\bra{\Psi}}_{i,j}  \right\} \right) = 1$. The results below  Table \ref{Tab:PauliKD} hold for every version of the $k=2$  KD distribution, where $ \{ \ket{a_i} \} $ is one Pauli basis, $ \{ \ket{f_j} \} $ is another Pauli basis, and $\ket{\Psi}$ is an eigenstate of the third Pauli operator. This conclusion can be checked directly.

\end{exmp}

\begin{exmp}[Nonclassical KD distribution that violates $2N_A+2N_F > 3d+n_\parallel-3\bar{n}_\parallel$] \label{Ex:NonClKDClassicalIneq} 

Satisfying  $2N_A+2N_F > 3d+n_\parallel-3\bar{n}_\parallel$ suffices to guarantee a nonclassical KD distribution. But it is not necessary, as we demonstrate here.  Consider a two-qubit system.  We choose  $\hat{A}$ and $\hat{F}$ such that  $\{ \ket{a_i} \} = \{ \ket{0}\ket{0}, \ket{0}\ket{1}, \ket{1}\ket{0}, \ket{1}\ket{1} \}$ and $\{ \ket{f_j} \} = \{ \ket{0}\ket{+}, \ket{0}\ket{-}, \ket{1}\ket{+}, \ket{1}\ket{-} \}$. 
We set $\hat{\rho} = \ket{\Psi} \bra{\Psi}$, where $\ket{\Psi} = \left( \ket{0}\ket{0}+2\ket{0}\ket{1} \right)/\sqrt{5}$. These choices imply $N_A=N_F=2$, $ d=4$, and $n_\parallel =  \bar{n}_\parallel=0$. Hence the inequality above is violated: $8 \ngtr 12$. Nonetheless, the KD distribution is nonclassical (Table \ref{Tab:NonClKDClassicalIneq}).

\begin{table}[h]
\caption{\label{Tab:NonClKDClassicalIneq}%
The KD distribution of Ex. \ref{Ex:NonClKDClassicalIneq}.
}
\begin{center}
\begin{tabular}{c|cccc}
\textrm{\backslashbox{$\ket{f_j}$}{$\ket{a_i}$} } &
\textrm{$\ket{0}\ket{0}$} &
\textrm{$\ket{0}\ket{1}$}&
\textrm{$\ket{1}\ket{0}$}&
\textrm{$\ket{1}\ket{1}$}\\
\hline \hline 
$\ket{0}\ket{+}$ & $\frac{3}{10}$ & $\frac{3}{5}$ & $0$ & $0$ \\
$\ket{0}\ket{-}$ & $-\frac{1}{10}$ & $\frac{1}{5}$ & $0$ & $0$
 \\
$\ket{1}\ket{+}$ & $0$ & $0$ & $0$& $0$
 \\
$\ket{1}\ket{-}$ & $0$ & $0$ & $0$ & $0$
\end{tabular}
\end{center}
\end{table}

\end{exmp}

\section{Proof of Thm. \ref{Th:NClKD}}
	 \label{App:Theo1Proof}

For convenience, we first assume that no $\ket{a_i}$ and $\ket{f_j}$ are parallel: $n_{\parallel}=\bar{n}_{\parallel}=0$.  Then, we generalize.

Assume that the KD distribution is classical: $ q^{\hat{\rho}}_{i,j}\in \mathbb{R}_{\geq 0}$ for all $i,j$. Without changing the quasiprobabilities or the observables, we can redefine the vectors through 
$ \ket{a_i} \mapsto e^{i \alpha_i} \ket{a_i}$  and 
$ \ket{f_j} \mapsto e^{i \phi_j} \ket{f_j}  $.
We choose the $\alpha_i, \phi_j  \in \mathbb{R}$ 
such that $\braket{a_i|\Psi} \braket{\Psi|f_j}\in \mathbb{R}_{\geq 0}$.  By assumption, $ \braket{f_j|a_i} \braket{a_i|\Psi} \braket{\Psi|f_j}\in \mathbb{R}_{\geq 0}$. Hence, for each $i$ and $j$, $\braket{a_i|f_j} \in \mathbb{R}_{\geq 0}$, or $\braket{a_i|\Psi}=0$, or $\braket{\Psi|f_j}=0$. Let $\hat{U}$ denote the unitary operator
that rotates $\mathcal{V}_A$ into $\mathcal{V}_F$. 
$\hat{U}$ is represented, relative to $\mathcal{V}_F$,
by the matrix with elements
$\hat{U}_{i,j}= \braket{a_i|f_j} $. $d-N_A$ vectors in $\mathcal{V}_A$, and $d-N_F$ vectors in $\mathcal{V}_F$, are orthogonal to $\ket{\Psi}$. Hence, at most $d-N_A$ rows and $d-N_F$ columns of $\hat{U}$ contain negative or nonreal values.

\begin{figure}
\includegraphics[scale=0.147]{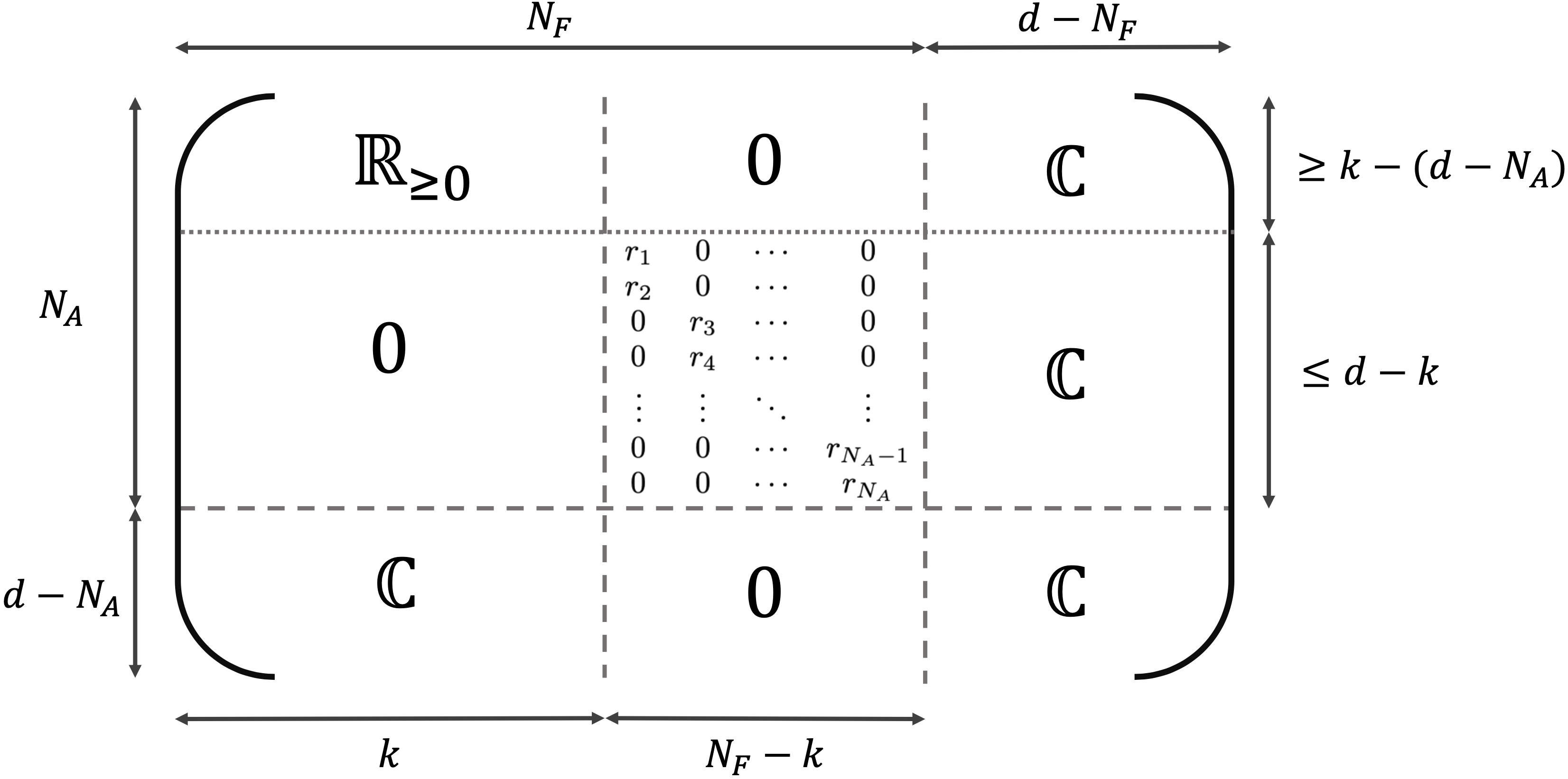}
\caption{Unitary matrix with entries $\hat{U}_{i,j} = \braket{a_i | f_j}$. 
The dashed vertical lines divide the columns into ``left,'' ``middle,'' and ``right'' sets. 
The dashed horizontal line divides the rows into ``top'' and ``bottom'' sets. The vectors $\ket{a_i}$ and $\ket{f_j}$ are ordered such that any nonreal or negative $\hat{U}_{i,j}$ appear in the bottom rows or rightmost columns. }
\label{fig:UnitaryMatrix}
\end{figure}

Let us order $\mathcal{V}_{A}$ and $\mathcal{V}_{F}$ so that 
the top left-hand $N_A$-by-$N_F$
block contains only non-negative real entries (Fig. \ref{fig:UnitaryMatrix}). The top $N_A$ entries of each column $j$ form a ``top vector’’  $\mathbf{f}_j^{\rm t} \in \mathbb{R}_{\geq 0}^{N_A}$. 
The bottom $d - N_A$ entries of column $j$ form a ``bottom vector’’  $\mathbf{f}_j^{\rm b} \in \mathbb{C}^{d-N_A}$. We label columns 1 to $k$ ``left,'' columns $k+1$ to $N_F$ ``middle,'' and columns $N_F+1$ to $d$ ``right." 
 
For $j = 1, 2, \ldots, N_F$, all elements of each $\mathbf{f}_j^{\rm t}$ are non-negative reals.
Hence $(\mathbf{f}_\ell^{\rm t})^{\top}  \mathbf{f}_m^{\rm t} \geq 0$ for all 
$\ell,  m \in\{1,\ldots,N_F\}$. 
Therefore, for the columns of $\hat{U}$ to be orthogonal, 
$(\mathbf{f}_\ell^{\rm b})^{\dagger} \mathbf{f}_m^{\rm b}\leq 0$ must hold for all 
$\ell, m  \in  \{1,\ldots,N_F\}$ for which $\ell \neq m$.  
This inner-product constraint implies the following lemma.
\newtheorem{lemma}{Lemma}
\begin{lemma}
\label{Lem:InProds}
At most $ 2(d-N_A)$ of the 
$N_F$ left and middle
bottom vectors are nonzero.
\end{lemma}
\textit{Proof of Lem. \ref{Lem:InProds}:}  
Here, we  bound the maximum number of nonzero bottom vectors whose pairwise products are $\leq 0$. Let $\mathcal{S} = \{ \mathbf{s}_j \}$ denote a set of nonzero vectors in $\mathbb{C}^n$ whose pairwise inner products are $\leq 0$. We use an orthonormal basis in terms of which $\mathbf{s}_1 \to (s_1,0,\ldots,0)^{\top}$ and $s_1 >0$. Every other vector $ \mathbf{s}_j \in \mathcal{S} \setminus \{ \mathbf{s}_1 \}$ is represented by a column with first element $\leq 0$. Hence, for these other $\mathbf{s}_j$ to have inner products $\leq 0$, the vectors formed from their last $n-1$ entries must all have inner products $\leq 0$. At most one of these shorter vectors can be the null vector $\mathbf{0}$. So all the others are nonzero vectors in $\mathbb{C}^{n-1}$ whose pairwise inner products are $\leq 0$. The relevant vectors space's dimensionality has decreased to $n-1$. Proceeding from $n$ to $n-1$, we have ``lost'' at most two vectors, $\mathbf{s}_1 \to (s_1,0,\ldots,0)^{\top}$ and $\mathbf{s}_2 \to (-s_2,0,\ldots,0)^{\top}$, where $s_1,s_2 > 0$. By induction, $\mathcal{S}$ can have at most $2n$ vectors. In the proof of Thm. \ref{Th:NClKD}, $n=d-N_A$. Consequently, $\leq 2(d-N_A)$ of the 
$N_F$ left and middle
bottom vectors are nonzero.$\square$

Lemma \ref{Lem:InProds} ensures that if $k$ denotes the number of nonzero elements of $\{\mathbf{f}_1^{\rm b},\ldots,\mathbf{f}_{N_F}^{\rm b}\}$, then
\begin{align}
\label{Eq:bottomvectorbound}
k\leq 2(d-N_A).
\end{align}
Let us order the columns of $\hat{U}$ so that the $k$ nonzero bottom vectors occupy columns 1 to $k$, while $\mathbf{f}_{k+1}^{\rm b}=\mathbf{f}_{k+2}^{\rm b}= \ldots =\mathbf{f}_{N_F}^{\rm b}= \mathbf{0}.$
(Fig. \ref{fig:UnitaryMatrix}).

Columns 1 to $k$ (the left columns) are linearly independent.
Therefore, the collection of columns contains nonzero entries in $\geq k$ rows. 
Up to $d - N_A$ of those rows can be in the bottom vectors
(which contain exactly $d - N_A$ rows).
The left top vectors make up the difference, having
nonzero entries in $\geq k-(d-N_A)$ rows. 
The middle top vectors must contain only 0s in these rows, 
since they are orthogonal to the left top vectors.\footnote{
The middle columns are orthogonal to the left columns.
The middle bottom columns' being $\mathbf{0}$s forces  the middle top vectors to be orthogonal to the left top vectors.}
Let us order the rows of $\hat{U}$ such that 
the middle top vectors'
uppermost $\geq k-(d-N_A)$ entries  
are 0s (Fig. \ref{fig:UnitaryMatrix}). 
Only the middle top vectors' lower 
$\leq  N_A-[k-(d-N_A)] = d-k$ entries can be nonzero. 
By assumption, no $\ket{a_i}$ is parallel to any $\ket{f_j}$.
So each middle top vector has $\geq 2$ nonzero entries $\langle a_i | f_j \rangle$. 
But the middle top  vectors are mutually orthogonal, 
and all their entries $\geq 0$.
So no two middle top vectors can have nonzero elements in the same row. 
Therefore, $2(N_F-k)\leq d-k$. We bound $k$ with Ineq. \ref{Eq:bottomvectorbound} and rearrange:
\begin{equation} 
\label{Eq:ReducedBound}
2N_A+2N_F  \leq  3d.
\end{equation}

Finally, we extend  $2N_F+2N_A\leq 3d$ [Ineq. \eqref{Eq:ReducedBound}]   to  scenarios in which $\bar{n}_{\parallel}\neq 0 $ or $n_{\parallel}\neq 0 $, completing the proof of Thm. \ref{Th:NClKD}. We first remove any pairs $( \ket{a_i}$, $\ket{f_j} )$ of parallel vectors from $\mathcal{V}_A$ and $\mathcal{V}_F$. Consider the subspace $\mathcal{H}^{\prime}$ spanned by the remaining basis vectors. Let $d^{\prime} \equiv \dim (\mathcal{H}^{\prime})$. 
Define $N_A^{\prime}$ as the number of $\ket{a_i}$
that have nonzero overlaps with $\ket{\Psi}$,
and define $N_F^{\prime}$ analogously.
Denote by $\ket{\Psi^{\prime}}$ the projection of  $\ket{\Psi}$  onto $\mathcal{H}^{\prime}$. Inequality \eqref{Eq:ReducedBound} can be rederived for this  reduced subspace: $2N_F^{\prime}+2N_A^{\prime}\leq 3d^{\prime}$.
(If $\ket{\Psi^{\prime}}=\bm{0}$, then $N_A^{\prime}=N_F^{\prime}=0$, so the inequality still holds.) Substituting in from $N_A^{\prime}+n_\parallel=N_A$, $N_F^{\prime}+n_\parallel=N_F$, and $d^{\prime}+n_{\parallel}+\bar{n}_{\parallel}=d$  leads to   $2N_F+2N_A\leq 3d+n_\parallel-3\bar{n}_\parallel $.

We derived this inequality assuming a classical KD distribution.
A violation of the inequality implies nonclassicality.$\square$

    \section{Properties of the imaginary and real components of the KD distribution} \label{App:ReImKD}
    
Consider an experiment that involves eigenbases $\{ \ket{a_i} \}$ and $\{ \ket{f_j} \}$ or, equivalently, nondegenerate operators $\hat{A}$ and $\hat{F}$. One might want to construct a KD distribution $\{ q^{\hat{\rho}}_{i,j} \}$ that has, or that lacks, KD nonclassicality by picking a suitable $\hat{\rho}$. Furthermore, one might want specific quasiprobabilities $q^{\hat{\rho}}_{i,j}$ to have negative or nonreal nonclassicality. We  provide useful results for tailoring  $\hat{\rho}$.
    
As in part of the main text, we assume that $\hat{\rho}$ is pure: $\hat{\rho} = \ket{\Psi}\bra{\Psi}$. 
The imaginary part of $q^{\hat{\rho}}_{i,j}$ decomposes  as
\begin{align}
\Im \left[ q^{\hat{\rho}}_{i,j} \right] = & \frac{1}{2i} \left[ q^{\hat{\rho}}_{i,j} -  \left( q^{\hat{\rho}}_{i,j} \right)^{*} \right] = \frac{1}{2} \mathrm{Tr} \left[ \hat{H}_{i,j}  \hat{\rho}  \right] ,
\end{align}
where $\hat{H}_{i,j} \equiv i \hat{\Pi}^a_i \hat{\Pi}^f_j - i  \hat{\Pi}^f_j \hat{\Pi}^a_i $.  If $|\braket{a_i | f_j}| \neq 0,1$, then $-i \hat{H}_{i,j}$ is the antisymmetric product of two noncommuting $\mathrm{rank}$-$1$ projectors. Under this condition, $ \hat{H}_{i,j}$ also has two  eigenvalues, $h_{i,j}^{(\pm)} = \pm |\braket{a_i | f_j}|\sqrt{1- |\braket{a_i | f_j}|^2} \neq 0$, with respective eigenvectors
\begin{align}
\ket{h_{i,j}^{(\pm)}} =   \frac{1}{\sqrt{2}} \Big[ \Big( \mp 1 - i \frac{|\braket{a_i | f_j}|}{\sqrt{1 - |\braket{a_i | f_j}|^2}} \Big) e^{i \mathrm{Arg}(\braket{a_i | f_j})} \ket{a_i} + i \frac{1}{\sqrt{1 - |\braket{a_i | f_j}|^2}}  \ket{f_j} 
\Big] .
\end{align}

The real part of $q^{\hat{\rho}}_{i,j}$ can be written as
\begin{equation}
\label{Eq:ReQPD}
\Re \big[ q^{\hat{\rho}}_{i,j} \big] = \frac{1}{2} \mathrm{Tr} \big[ q^{\hat{\rho}}_{i,j} +  \left( q^{\hat{\rho}}_{i,j} \right)^{*} \big] \equiv \frac{1}{2} \mathrm{Tr} \big[ \hat{G}_{i,j}  \hat{\rho}  \big] ,
\end{equation}
where $\hat{G}_{i,j}  \equiv \hat{\Pi}^a_i \hat{\Pi}^f_j +  \hat{\Pi}^f_j \hat{\Pi}^a_i$.  If $|\braket{a_i | f_j}| \neq 0$, then $\hat{G}_{i,j}$ is the symmetric product of two noncommuting $\mathrm{rank}$-$1$ projectors. Under this condition,  $\hat{G}_{i,j}$ also has two eigenvalues, $g_{i,j}^{(\pm)} = |\braket{a_i | f_j}|(|\braket{a_i | f_j}| \pm 1) \neq 0$, with corresponding eigenvectors
\begin{align}
\ket{g_{i,j}^{(\pm)}} =\frac{1}{\sqrt{2}}  \Big( \ket{f_j}   \pm e^{i \mathrm{Arg}(\braket{a_i | f_j})} \ket{a_i}  \Big) .
\end{align}
$g_{i,j}^{(+)}$ and $g_{i,j}^{(-)}$ are  positive and negative, respectively.  This result is consistent with the appendix in Ref. \cite{Hartle04}. There, Hartle demonstrates the existence of a state for a which a KD distribution is nonclassical, if the two projectors fail to commute.

Given the eigenvalues $h_{i,j}^{(\pm)}$ and $g_{i,j}^{(\pm)}$, and the eigenvectors $\ket{h_{i,j}^{(\pm)}} $ and $\ket{g_{i,j}^{(\pm)}} $, one can tailor $\ket{\Psi}$ such that a  quasiprobability $q^{\hat{\rho}}_{i,j}$  has a negative real component, or an imaginary component, of a certain magnitude.

    \section{Extension to restricted information, or coarse-grained KD distributions} \label{App:CoarseKDExt}

 $\hat{A}$ can be degenerate, as can $\hat{F}$. Regardless, $\hat{A}$ eigendecomposes as $\hat{A} = \sum_l A_l \hat{A}_l$, where  $ \hat{A}_l \equiv \sum_{i \, : \, \ket{a_i} \in \mathcal{A}_l} \ket{a_i}\bra{a_i}$ and $\mathcal{A}_l$ is the eigensubspace associated with the eigenvalue $A_l$. Similarly, $\hat{F} = \sum_k F_k \hat{F}_k$, where  $ \hat{F}_k \equiv \sum_{j \, : \, \ket{f_j} \in \mathcal{F}_k} \ket{f_j}\bra{f_j}$ and $\mathcal{F}_k$ is eigensubspace associated with the eigenvalue $F_k$. If any $\hat{F}_k$ ($ \hat{A}_l $) has rank $>1$,  $\hat{F}_k$  ($ \hat{A}_l $)  has  nonequivalent eigenbases. Consequently, $ \{ q^{\hat{\rho}}_{i,j} \}$ is generally not  unique for a fixed $\hat{\rho}$. This degeneracy problem arises  in, e.g., studies of quantum scrambling:  $\hat{A}$ and $\hat{F}$ manifest as local observables of a many-body system and so are degenerate  \cite{Yunger18, Yunger18-2, Yunger19, Razieh19}.   We therefore define a \emph{coarse-grained KD quasiprobability distribution} by  marginalizing $ \{ q^{\hat{\rho}}_{i,j} \} $ over the degeneracies:
\begin{align}
\label{eq_Coarse_Q}
\left\{ \mathcal{Q}^{\hat{\rho}}_{l,k} \right\} := \left\{ \sum_{ \substack{ i \, : \, \ket{a_i} \in \mathcal{A}_l \\ j \, : \, \ket{f_j} \in \mathcal{F}_k } } \braket{f_j | a_i} \bra{a_i} \hat{\rho} \ket{ f_j} \right\} = \left\{ \mathrm{Tr} \left( \hat{F}_k \hat{A}_l \hat{\rho}  \right) \right\}.
\end{align}
The projectors $\hat{F}_k$ and $\hat{A}_l$ are unique. So, for a given $\hat{\rho}$, the quasiprobabilities $ \mathcal{Q}^{\hat{\rho}}_{l,k}  $ are unique.

We now prove a theorem analogous to Thm. \ref{Th:NClKD} for the coarse-grained distribution, providing a necessary condition for $\{ \mathcal{Q}^{\hat{\rho}}_{l,k}  \} $ to be classical when $\hat{\rho} = \ket{\Psi}\bra{\Psi}$ is pure. 
In analogy with Eq. \eqref{Eq:Na}, we define as $\tilde{N}_A$ 
the number of $\hat{A}$ eigenspaces onto which
$\ket{\Psi}$ has nonzero projections.
In analogy with Eq.~\eqref{Eq:Nf}, we define $\tilde{N}_F$ similarly:
\begin{align}
\tilde{N}_A & \equiv || \{ l : \hat{A}_l\ket{\Psi} \neq 0 \} || , \; \; \mathrm{and} \\
\tilde{N}_F & \equiv || \{ k : \hat{F}_k\ket{\Psi} \neq 0 \} || .
\end{align}
In analogy with previous definitions, 
we denote by $\tilde{n}_\parallel$ 
(respectively, $\tilde{\bar{n}}_\parallel$) 
the number of   $\hat{A}_l\ket{\Psi}$ that are 
(i) parallel to some $ \hat{F}_k\ket{\Psi}$  and 
(ii) nonorthogonal (respectively, orthogonal) to $\ket{\Psi}$. 
This background informs the following theorem, which resembles Thm. \ref{Th:NClKD}.

\begin{theorem}[Sufficient conditions for coarse-grained Kirkwood-Dirac nonclassicality]
	\label{Th:NClExtKD}
	Suppose that $\hat{\rho} = \ket{\Psi} \bra{\Psi}$ is pure. If $2\tilde{N}_F+2\tilde{N}_A>3d+\tilde{n}_\parallel-3\tilde{\bar{n}}_\parallel$, the coarse-grained KD distribution is nonclassical.
\end{theorem}
\textit{Proof:} 
As in the proof of Thm. \ref{Th:NClKD}, we begin by assuming that the KD distribution is classical: $\mathcal{Q}^{\hat{\rho}}_{l,k}\geq 0$ for all $l,k$. We assume that $\tilde{n}_\parallel=\tilde{\bar{n}}_\parallel=0$; then,  we generalize.

Define the $\tilde{N}_A$ nonzero projections $\ket{a_l^{\Psi}}\equiv\hat{A}_l\ket{\Psi}/||\hat{A}_l\ket{\Psi}||$ and the $\tilde{N}_F$ nonzero projections $\ket{f_k^{\Psi}}\equiv\hat{F}_k\ket{\Psi}/||\hat{F}_k\ket{\Psi}||$. 
By appending vectors to the sets 
$\{\ket{a_l^{\Psi}}\}$ and $\{\ket{f_k^{\Psi}}\}$,
we can form orthonormal bases $\mathcal{B}_A$ and $\mathcal{B}_F$. 
By the sets' definitions, $\ket{\Psi}\in $ span$ \{\ket{a_l^{\Psi}}\}$, and $\ket{\Psi}\in $ span$ \{\ket{f_k^{\Psi}}\}$.
Therefore, the appended vectors are orthogonal to $\ket{\Psi}$.
Since $\mathcal{Q}^{\hat{\rho}}_{l,k} 
= \mathrm{Tr} \left( \hat{F}_k \hat{A}_l \hat{\rho}  \right)
= \bra{\Psi}\hat{F}_k \hat{A}_l\ket{\Psi}
= \braket{f_k^{\Psi}|a_l^{\Psi}}   \times
||\hat{A}_l\ket{\Psi}||   \times 
||\hat{F}_k\ket{\Psi}||$, 
the condition $\mathcal{Q}^{\hat{\rho}}_{l,k}\geq 0$ implies that $\braket{f_k^{\Psi}|a_l^{\Psi}}\geq 0$. 
Therefore, any nonclassical quasiprobabilities contain
vectors appended to the bases $\mathcal{B}_A$ and $\mathcal{B}_F$.
But the appended basis elements are orthogonal to $\ket{\Psi}$
and so appear only in zero-valued quasiprobabilities. 
Therefore, $\mathcal{B}_A$ and $\mathcal{B}_F$ define a classical non-coarse-grained KD distribution for $\hat{\rho}$. 
Let this non-coarse-grained KD distribution's 
$N_A,N_F,n_\parallel$ and $\bar{n}_\parallel$ 
be defined as in the proof of Thm. \ref{Th:NClKD}.
By Thm.  \ref{Th:NClKD}, $2N_A+2N_F\leq 3d+n_\parallel-3\bar{n}_\parallel$.
Since we extended the bases with vectors orthogonal to $\ket{\Psi}$, 
$N_A=\tilde{N}_A$, $N_F=\tilde{N}_F$, and 
$  n_\parallel  =  \tilde{n}_\parallel =0$. 
Therefore, $2\tilde{N}_A+2\tilde{N}_F\leq 3d-\bar{n}_\parallel \leq 3d$. 
The generalization to $\tilde{n}_\parallel\neq 0$ or $\tilde{\bar{n}}_\parallel\neq 0$ proceeds as in App.~\ref{App:ReducedBoundExt}. Therefore, every classical coarse-grained KD distribution satisfies 
$2\tilde{N}_F+2\tilde{N}_A \leq 3d+\tilde{n}_\parallel-3\tilde{\bar{n}}_\parallel$. Violating this inequality  suffices for the coarse-grained distribution to be nonclassical. $\square$

An analog of Cor. \ref{Cor:KD0} follows.

\begin{Cor}
\label{Th:KD0Ext}
	Suppose that at least one of $\hat{A}$ and $\hat{F}$ is nondegenerate, while the other is not completely degenerate. If the KD distribution lacks zero-valued quasiprobabilities, $\mathcal{Q}^{\hat{\rho}}_{l,k}$ is nonclassical.
\end{Cor}

\textit{Proof:} 
Suppose that all the $\mathcal{Q}^{\hat{\rho}}_{l,k}$ are nonzero. 
Without loss of generality, 
assume that $\hat{A}$ is nondegenerate.
$\hat{F}$ is not completely degenerate, 
so its eigendecomposition contains 
at least two distinct projectors, $\hat{F}_1$ and $\hat{F}_2$. 
Since the $\mathcal{Q}^{\hat{\rho}}_{l,k}$ are nonzero, 
$\hat{F}_1\ket{\Psi}$ and $\hat{F}_2\ket{\Psi}$ are nonzero,
by Eq.~\eqref{eq_Coarse_Q}.
Therefore, there exist at least two vectors,
$\ket{f_1^\Psi}$ and $\ket{f_2^\Psi}$, 
as defined in the proof of Thm. \ref{Th:NClExtKD}. 

The rest of the proof is a proof by contradiction.
Suppose that $ \{ \mathcal{Q}^{\hat{\rho}}_{l,k} \}$ 
is classical. If it lacks zero-valued quasiprobabilities, then 
$\braket{f_k^{\Psi}|a_l}\in \mathbb{R}_{> 0}$ 
for every $l$ and $k$. 
By the $\hat{F}$ eigenspaces' orthogonality, 
\begin{align}
0&=\braket{f_1^\Psi|f_2^\Psi}
=\sum_l \braket{f_1^\Psi|a_l}\braket{a_l|f_2^\Psi}
>0 . 
\end{align} 
The final inequality follows because 
$\braket{f_1^\Psi|a_l}, \braket{a_l|f_2^\Psi} > 0$
for each $l$.
Implying the contradiction $0 > 0$,
the assumption of the distribution's classicality is false. $\square$\\

Let us briefly discuss the case, 
consistent with the assumptions of Cor. \ref{Th:KD0Ext}, 
in which $\hat{F}$ is degenerate and $\hat{A}$ is not (or \textit{vice versa}). 
Coarse-graining over one index suffices
to define a unique KD distribution distribution:
\begin{align}
\left\{ Q^{\hat{\rho}}_{i,k} \right\} 
:= \left\{  \sum_{j \, : \, \ket{f_j} \in \mathcal{F}_k} \braket{f_j | a_i} \bra{a_i} \hat{\rho} \ket{ f_j} \right\} 
= \left\{ \mathrm{Tr} \left( \hat{F}_k \ket{a_i}\bra{a_i} \hat{\rho}  \right) \right\} .
\end{align}
 Such a distribution has been used, for example, in  postselected quantum metrology.  In Ref. \cite{ArvShukur19-2}, $ \hat{F} = 0 \times \sum_{j \, : \, \ket{f_j} \in \mathcal{F}_0} \ket{f_j}\bra{f_j} +  1 \times \sum_{j^{\prime} \, : \, \ket{f_j^{\prime}} \in \mathcal{F}_1} \ket{f_j^{\prime}}\bra{f_j^{\prime}} $ is an observable whose measured value determines whether a quantum state should be discarded or  funnelled to further processing.  If the coarse-grained KD distribution contains  negative values, a metrological protocol may provide a nonclassical advantage. Further properties of $Q^{\hat{\rho}}_{i,k}$ are proved below.

    \subsection{Properties of the imaginary and real components of the coarse-grained KD distribution} \label{App:ReImMargKD}

Here, we extend the results of App. \ref{App:ReImKD} to $ \{ Q^{\hat{\rho}}_{i,k} \} $.  Suppose that $\hat{\rho}$ is pure: $\hat{\rho} = \ket{\Psi}\bra{\Psi}$.  
The imaginary part of $Q^{\hat{\rho}}_{i,k}$ decomposes  as 
\begin{align}
\Im \left[ Q^{\hat{\rho}}_{i,k} \right] =  & \frac{1}{2i} \left[ Q^{\hat{\rho}}_{i,k}  -  \left( Q^{\hat{\rho}}_{i,k}  \right)^{*} \right] = \frac{1}{2} \mathrm{Tr} \left[ \hat{R}_{i,k}  \hat{\rho}  \right] ,
\end{align}
where $\hat{R}_{i,k} \equiv  i \hat{\Pi}^a_i \hat{F}_k- i  \hat{F}_k \hat{\Pi}^a_i$. If $p^a_F \equiv \mathrm{Tr}\left( \hat{\Pi}^a_i \hat{F}_k \right) \neq 0,1 $, then $\hat{R}_{i,k}$ has two nonzero eigenvalues, $r_{i,k}^{(\pm)} = \pm \sqrt{p^a_F - \left( p^a_F \right)^2 }$. The eigenvectors are
\begin{equation}
\ket{r_{i,k}^{(\pm)}} =   \frac{1}{\sqrt{2}} \left[ \left( \mp \frac{1}{\sqrt{p^a_F}} + i \frac{1}{\sqrt{1 - p^a_F}} \right) \hat{F}_k \ket{a_i} - i \frac{1}{\sqrt{1 -p^a_F}}  \ket{a_i} 
\right] .
\end{equation}

Similarly, the real part of $Q^{\hat{\rho}}_{i,k}$ can be expressed as
\begin{align}
\Re \left[ Q^{\hat{\rho}}_{i,k} \right] =  & \frac{1}{2} \left[ Q^{\hat{\rho}}_{i,k}  +  \left( Q^{\hat{\rho}}_{i,k}  \right)^{*} \right] = \frac{1}{2} \mathrm{Tr} \left[ \hat{S}_{i,k}  \hat{\rho}  \right] ,
\end{align}
where $\hat{S}_{i,k} \equiv  \hat{\Pi}^a_i \hat{F}_k + \hat{F}_k \hat{\Pi}^a_i$. If $p^a_F \neq 0,1 $, then $\hat{S}_{i,k}$ has two eigenvalues, $s_{i,k}^{(\pm)} = p^a_F \pm \sqrt{p^a_F }$. The eigenvectors are
 \begin{equation}
\ket{s_{i,k}^{(\pm)}} =   \frac{1}{\sqrt{2}} \left[ \ket{a_i} \pm \frac{1}{\sqrt{p^a_F}} \hat{F}_k  \ket{a_i} 
\right] .
\end{equation}

\section{Proof of Thm. \ref{Th:MaxNCli}}
	 \label{App:Theo2Proof}

Here, we upper-bound  $\mathcal{N} \left( \{ q^{\hat{\rho}}_{i_1, \ldots, i_k} \} \right)$,
proving Thm. \ref{Th:MaxNCli}. 
First, we restrict our attention pure states $\hat{\rho}=\ket{\Psi}\bra{\Psi}$. We prove that $\mathcal{N} \left( \{ q^{\ket{\Psi}\bra{\Psi}}_{i_1,\ldots, i_k} \} \right)$ maximizes when each of its inner products has magnitude  $1/\sqrt{d}$. Thus, if $\mathcal{N} \left( \{ q^{\ket{\Psi}\bra{\Psi}}_{i_1,\ldots, i_k} \} \right)$ is maximized, then $|\braket{a^{(1)}_{i_1}|\Psi}|=|\braket{a^{(k)}_{i_k}|\Psi}|=\frac{1}{\sqrt{d}}$ for all $i_1,i_k$. Every $\hat{\rho}$ equals a convex sum of pure states $\hat{\rho}_n$. By the triangle inequality, $\mathcal{N} \left( \{ q^{\hat{\rho}}_{i_1, \ldots, i_k} \} \right)$ is upper-bounded by a convex sum of the $\mathcal{N} \left( \{ q^{\hat{\rho}_n}_{i_1, \ldots, i_k} \} \right)$. Therefore, at any maximum of $\mathcal{N} \left( \{ q^{\hat{\rho}}_{i_1, \ldots, i_k} \} \right)$, $\hat{\rho}$ is a linear combination of pure states,  each of which maximizes $\mathcal{N} \left( \{ q^{\hat{\rho}}_{i_1, \ldots, i_k} \} \right)$. We finish the proof by showing that no such mixed state maximizes $\mathcal{N} \left( \{ q^{\hat{\rho}}_{i_1, \ldots, i_k} \} \right)  $. Hence, only pure states that are unbiased with respect to $\hat{A}_1$ and $\hat{A}_k$ eigenbases, as described above, maximize $\mathcal{N} \left( \{ q^{\hat{\rho}}_{i_1, \ldots, i_k} \} \right) $.

Our proof requires the following lemma:
\newtheorem{lemma2}{Lemma}
\begin{lemma2}
\label{Lem:One}
Let $\{\ket{i}\}_{i=1}^d$ be an orthonormal basis for a $d$-dimensional Hilbert space $\mathcal{H}$.
The unit vector $\ket{\psi}\in \mathcal{H}$ satisfies 
$\sum_{i=1}^d |\braket{i|\psi}|\leq \sqrt{d}$.
The bound is saturated if and only if  
$|\braket{i|\psi}|=\frac{1}{\sqrt{d}}$ for every $i$.
\end{lemma2}

\textit{Proof:}
By Jensen's inequality, 
\begin{align}
\label{eq_Jensen}
\left( \sum_{i=1}^d |\braket{i|\psi}| \right)^2 \leq d \sum_{i=1}^d |\braket{i|\psi}|^2  = d .
\end{align}
Comparing the first and third expressions, 
we conclude that
\begin{align}
   \label{eq_Lem_One}
   \sum_{i=1}^d |\braket{i|\psi}| \leq \sqrt{d} .
\end{align}
Jensen's inequality is saturated if and only if
the terms in the first sum in~\eqref{eq_Jensen} equal each other,
as can be inferred from the geometric proof of Jensen's inequality.
Consequently, Ineq.~\eqref{eq_Lem_One} is saturated 
if and only if $|\braket{i|\psi}| = 1/\sqrt{d}$. $\square$

To upper-bound $\mathcal{N} \left( \{ q^{\hat{\rho}}_{i_1, \ldots, i_k} \} \right)$, we assume that $\rho=\ket{\Psi}\bra{\Psi}$ is pure. By Eqs. \eqref{Eq:KDExt} and \eqref{Eq:AggNonClas},
\begin{align}
\label{Eq:AppNonClas}
\mathcal{N} \left( \left\{ q^{\ket{\Psi}\bra{\Psi}}_{i_1, \ldots, i_k} 
\right\}  \right)  
= -1+\sum_{i_1, \ldots, i_k} |\braket{a^{(1)}_{i_1}|a^{(2)}_{i_2}}\times \ldots \times \braket{a^{(k)}_{i_k}|\Psi}\braket{\Psi |a^{(1)}_{i_1}}|,
\end{align} where $\{ \ket{a_{i_n}^{(n)} }\}_{i_n=1}^d$ is an eigenbasis of Hermitian operator $A^{(n)}$.  [To simplify notation in this proof, we have labeled operators differently than in Eq. \eqref{Eq:KDExt}: Here, the $\bra{ a^{(k)}_{i_k} }$ acts on $\ket{\Psi}$.] We now show that the RHS of Eq. \eqref{Eq:AppNonClas} maximizes when the magnitude of all the inner products in $\mathcal{N} \left( \{ q^{\ket{\Psi}\bra{\Psi}}_{i_1,\ldots, i_k} \} \right)$ equal each other.

For a fixed value of $i_1$, 
\begin{align}
\sum_{i_2, \ldots, i_k} |\braket{a^{(1)}_{i_1}|a^{(2)}_{i_2}}\times \ldots \times \braket{a^{(k)}_{i_k}|\Psi}\braket{\Psi |a^{(1)}_{i_1}}|&=\sum_{i_2}\left(|\braket{a^{(1)}_{i_1}|a^{(2)}_{i_2}}|\times \sum_{i_3, \ldots, i_k} |\braket{a^{(2)}_{i_2}|a^{(3)}_{i_3}}\times \ldots \times \braket{a^{(k)}_{i_k}|\Psi}\braket{\Psi |a^{(1)}_{i_1}}|\right) 
\label{Eq:NonClasPrior}
\\
&\leq \sum_{i_2}|\braket{a^{(1)}_{i_1}|a^{(2)}_{i_2}}|\times \max_{i_2'}\sum_{i_3, \ldots, i_k} |\braket{a^{(2)}_{i_2'}|a^{(3)}_{i_3}}\times \ldots \times \braket{a^{(k)}_{i_k}|\Psi}\braket{\Psi |a^{(1)}_{i_1}}|
\label{Eq:NonClasFirstIneq}
\\
&\leq \sqrt{d} \times\max_{i_2'}\sum_{i_3, \ldots, i_k} |\braket{a^{(2)}_{i_2'}|a^{(3)}_{i_3}}\times \ldots \times \braket{a^{(k)}_{i_k}|\Psi}\braket{\Psi |a^{(1)}_{i_1}}|.
\label{Eq:NonClasSecondIneq}
\end{align}
Inequality \eqref{Eq:NonClasFirstIneq} follows because, if $x_j$ and $y_j$ are non-negative real numbers, then $\sum_j x_j y_j \leq \sum_j x_j \times \max_{j'} y_{j'}$. Inequality \eqref{Eq:NonClasSecondIneq} follows from Lemma \ref{Lem:One}. 
Proceeding from the left-hand side  of Eq. \eqref{Eq:NonClasPrior} to the RHS of \eqref{Eq:NonClasSecondIneq}, we (i) reduce the number of summed indices by $1$ and (ii) acquire a factor of $\sqrt{d}$. Let us iterate this step $k-3$ more times:
\begin{align}
\sum_{i_2, \ldots, i_k} |\braket{a^{(1)}_{i_1}|a^{(2)}_{i_2}}\times \ldots \times \braket{a^{(k)}_{i_k}|\Psi}\braket{\Psi |a^{(1)}_{i_1}}| &\leq (\sqrt{d})^2\times \max_{i_3'}\sum_{i_4, \ldots, i_k} |\braket{a^{(3)}_{i_3'}|a^{(4)}_{i_4}}\times \ldots \times \braket{a^{(k)}_{i_k}|\Psi}\braket{\Psi |a^{(1)}_{i_1}}|\\
&\leq \ldots  \\
&\leq (\sqrt{d})^{k-2}\times 
\max_{i_{k-1}'}\sum_{i_k}
|\braket{a_{i'_{k-1}}^{(k-1)}  | a_{i_k}^{(k)} }
\braket{a_{i_k}^{(k)} |\Psi}   \braket{\Psi| a^{(1)}_{i_1} }| . \label{Eq:NonClasMoreIneq}
\end{align}
Summing over $i_1$ yields
\begin{align}
\sum_{i_1, \ldots, i_k} |\braket{a^{(1)}_{i_1}|a^{(2)}_{i_2}}\times \ldots \times \braket{a^{(k)}_{i_k}|\Psi}\braket{\Psi |a^{(1)}_{i_1}}|&\leq (\sqrt{d})^{k-2}\times \max_{i_{k-1}'}\sum_{i_1,i_k}|\braket{a^{(k-1)}_{i_{k-1}'}|a_{i_k}}\braket{a^{(k)}_{i_k}|\Psi}\braket{\Psi|a^{(1)}_{i_1}}| \\
&= d^{\frac{k}{2}-1} \sum_{i_1} |\braket{\Psi|a^{(1)}_{i_1}}|\times \max_{i_{k-1}'}\sum_{i_k}|\braket{a^{(k-1)}_{i_{k-1}'}|a^{(k)}_{i_k}}\braket{a^{(k)}_{i_k}|\Psi}|\\
&\leq d^\frac{k-1}{2} \times 
\max_{i_{k-1}'}
\sum_{i_k}
|\braket{a^{(k-1)}_{i_{k-1}'}|a^{(k)}_{i_k}}| 
\times 
|\braket{a^{(k)}_{i_k}|\Psi}|
\label{Eq:NonClasThirdIneq}
\\
&\leq d^\frac{k-1}{2} \times 
\max_{i_{k-1}'} \sqrt{\sum_{i_k}
|\braket{a^{(k-1)}_{i_{k-1}'}|
a^{(k)}_{i_k}}|^2\times
\sum_{i_k'}   |\braket{a^{(k)}_{i_k'}|\Psi}|^2}
\label{Eq:NonClasFourthIneq}
\\
&=d^{\frac{1}{2}(k-1)} .
\end{align}
Inequality \eqref{Eq:NonClasThirdIneq} follows from Lemma \ref{Lem:One}. 
Inequality \eqref{Eq:NonClasFourthIneq} follows from 
the Cauchy-Schwarz inequality:
For vectors $\vec{u}, \vec{v} \in \mathbb{R}^n$,
denote the inner product by
$(\vec{u},  \vec{v})  =  \sum_{j = 1}^d u_j v_j$.
The Cauchy-Schwarz inequality implies that
$(\vec{u}, \vec{v} )^2
\leq  (\vec{u},  \vec{u})  (\vec{v},  \vec{v})$.
Let $\vec{u}  
= \left( | \langle a^{(k-1)}_{i'_{k-1}} | a^{(k)}_1 \rangle | ,  
           | \langle a^{(k-1)}_{i'_{k-1}} | a^{(k)}_2 \rangle | ,
           \ldots,
           | \langle a^{(k-1)}_{i'_{k-1}} | a^{(k)}_d \rangle |  \right)$ and
$\vec{v}
= \left( | \langle a^{(k)}_1 | \Psi \rangle | , 
           | \langle a^{(k)}_2 | \Psi \rangle | ,
            \ldots,
           | \langle a^{(k)}_d | \Psi  \rangle \right)$.
Square-rooting each side of the Cauchy-Schwarz inequality 
yields Ineq.~\eqref{Eq:NonClasFourthIneq}. Therefore, 
\begin{equation}
\label{Eq:NonClasIneqApp}
\mathcal{N} \left( \{ q^{\hat{\rho}}_{i_1, \ldots, i_k} \} \right) \leq d^{(k-1)/2}-1 .
\end{equation}

It is easy to see that, if all the inner products in $\{ q^{\hat{\rho}}_{i_1, \ldots, i_k} \}$ have magnitudes $1/\sqrt{d}$, Ineq.~\eqref{Eq:NonClasIneqApp} is saturated. This criterion is satisfied when two conditions hold simultaneously:
(i) $\hat{A}^{(i)}$ and $\hat{A}^{(i+1)}$ have mutually unbiased eigenbases for each $i = 1, 2, \ldots ,k-1$; and (ii) $|\braket{a^{(1)}_{i_1}|\Psi}|=|\braket{a^{(k)}_{i_k}|\Psi}|=\frac{1}{\sqrt{d}}$ for all $i_1,i_k$. 

These two conditions are not only sufficient, but also necessary for $\mathcal{N} \left( \{ q^{\hat{\rho}}_{i_1, \ldots, i_k} \} \right)$ to be maximized: Inequalities \eqref{Eq:NonClasFirstIneq}-\eqref{Eq:NonClasMoreIneq} are all saturated only if (i) holds. Inequalities \eqref{Eq:NonClasThirdIneq} and \eqref{Eq:NonClasFourthIneq} are saturated only if (ii) holds.

Therefore, if a (possibly mixed) state $\hat{\rho}$ maximizes $\mathcal{N} \left( \{ q^{\hat{\rho}}_{i_1, \ldots, i_k} \} \right)$, then $\hat{\rho}=\sum_n p_n \ket{\Psi_n}\bra{\Psi_n}$, where each $\ket{\Psi_n}$ maximizes $\mathcal{N} \left( \{ q^{\ket{\Psi_n}\bra{\Psi_n}}_{i_1, \ldots, i_k} \} \right)$. By the triangle inequality, $ |\braket{a^{(k)}_{i_k}|\hat{\rho}|a^{(1)}_{i_1}}|\leq \sum_n p_n |\braket{a^{(k)}_{i_k}|\Psi_n}\braket{\Psi_n |a^{(1)}_{i_1}}|$, with equality only if $\arg \left( \braket{a^{(k)}_{i_k}|\Psi_n}\braket{\Psi_n |a^{(1)}_{i_1}} \right)$ is independent of $n$. So, if $\hat{\rho}$ maximizes $\mathcal{N} \left( \{ q^{\hat{\rho}}_{i_1, \ldots, i_k} \} \right)$, then, for each $i_1$ and $i_k$, $\arg \left( \braket{a^{(k)}_{i_k}|\Psi_n}\braket{\Psi_n |a^{(1)}_{i_1}} \right)$ is independent of $n$. Thus, since $|\braket{a^{(k)}_{i_k}|\Psi_n}\braket{\Psi_n |a^{(1)}_{i_1}}|=\frac{1}{d}$ for all $n$, $i_1$, and $i_k$, $\braket{a^{(k)}_{i_k}|\Psi_n}\braket{\Psi_n |a^{(1)}_{i_1}}$ is independent of $n$ for every $i_1,i_k$. Therefore, $\ket{\Psi_n}\bra{\Psi_n}$ is independent of $n$, and so $\rho$ is a pure state, as claimed. 
$\square$

\section{Real MUBs used to maximize $\mathcal{N}^{\Re^-}$ }
	 \label{App:RealMUBs}

A Kirkwood-Dirac distribution achieves its maximal negativity when
$\mathcal{N}^{\Re^-}  =  \max \{ \mathcal{N} \}$.
Such a distribution can be constructed from a triplet of real MUBs.
For our purposes, a real MUB is an MUB whose vectors can be represented,
relative to some basis, as columns of real numbers.
We now reconcile that definition with the definition in the literature.

Real MUBs have been defined as MUBs for 
Hilbert spaces over $\mathbb{R}^m$,
for $m = 2, 3, \ldots$ \cite{Boykin05}.
In contrast, we focus on Hilbert spaces over $\mathbb{C}^m$.
But real MUBs can be imported into complex vector spaces, as follows.

Let $\{ B_1, B_2, \ldots, B_n \}$ denote a set of 
real MUBs for $\mathbb{R}^m$,
and let $B_j = \{ \ket{ b^\JParen_1 },  \ldots, \ket{ b^\JParen_m }  \}$.
Each vector in $\mathbb{R}^m$ exists in $\mathbb{C}^m$,
so each $\ket{ b^\JParen_k }$ exists in $\mathbb{C}^m$.
Consider any vector $\ket{\vecc}$ that exists in $\mathbb{C}^m$
but not in $\mathbb{R}^m$.
$\ket{\vecc}$ equals a linear combination,
weighted with complex coefficients, of $\mathbb{R}^m$ vectors.
Every $\mathbb{R}^m$ vector equals a linear combination 
of the $\ket{ b^\JParen_k }$.
Therefore, $\ket{\vecc}  \in  \mathbb{C}^m$ equals a linear combination
of the $\ket{ b^\JParen_k }$.
So each $B_j$ is a basis for $\mathbb{C}^m$,
so $\{ B_1, \ldots, B_n \}$ forms a set of MUBs in $\mathbb{C}^m$.

Let $\Basis$ denote any basis for $\mathbb{R}^m$.
Relative to $\Basis$, every $\ket{ b^\JParen_k }$ 
can be represented as a column of real numbers,
by the definition of $\mathbb{R}^m$.
$\Basis$ forms a basis also for $\mathbb{C}^m$,
by the preceding paragraph.
Therefore, every $\ket{ b^\JParen_k }$ can be represented,
relative to a basis $\Basis$ for $\mathbb{C}^m$,
as a column of real numbers.

\onecolumngrid
\bibliography{KDNegDraft}

\end{document}